# Hydro-Gravitational-Dynamics Interpretation of the Tadpole VV29 Merging Galaxy System: Dark-Matter-Halo-Planet Star-Cluster Wakes

Carl H. Gibson [1,2]

[1] University of California San Diego, La Jolla, CA 92093-0411, USA
[2] cgibson@ucsd.edu, http://sdcc3.ucsd.edu/~ir118

**Abstract:** Hubble Space telescope (HST) images of merging galaxy system VV29 reveal the 0.3 Mpc ($10^{22}$ m) baryonic-dark-matter (BDM) halo composed of primordial protoglobularstarcluster (PGC) clumps of planets. Star-cluster-wakes trace the merger by formation of stars from the planets. Aligned young globular star clusters (YGCs), star-wakes and dust-trails show the frictional, spiral passage of galaxy fragments VV29cdef in a long tail-like galaxy (VV29b) as the fragments merge on the accretion disk plane of the central spiral galaxy VV29a. The observations confirm the hydro-gravitational-dynamics (HGD) prediction of Gibson 1996 and quasar microlensing inference of Schild 1996: i. e., that the dark matter of galaxies is dominated by planets (PFPs) in million-solar-mass clumps. Globular star clusters (YGCs, OGCs, PGCs) preserve the density $\rho_0 \sim 10^{-17}$ kg m$^{-3}$ of the plasma epoch 30,000 years after the big bang when viscous supercluster-fragmentation began. Tadpole images show linear galaxy clusters reflecting turbulent vortex lines of protogalaxy fragmentation at the 0.003 Mpc Kolmogorov-Nomura ($L_N$) scale of the plasma before transition to gas. The halo size indicates strong diffusion of PGC primordial-planet-clumps from a cooling protogalaxy as its planets freeze.

## 1. Introduction

Modern fluid mechanical techniques of engineering, oceanography and atmospheric sciences have been applied to cosmological and astrophysical problems of gravitational structure formation in a series of studies [1-5] termed Hydro-Gravitational-Dynamics (HGD). Standard cold-dark-matter hierarchical-clustering (CDMHC) cosmological models neglect critically important HGD concepts such as viscosity[1], diffusion[2], turbulence[3], fossil turbulence[4] and fossil turbulence waves[5], and should be abandoned [6-10]. HGD predicts Planck scale big bang turbulent combustion [4] terminated by muon-viscosity-driven space inflation at the strong-force freeze-out temperature to produce fossil-temperature-turbulence remnants [3]. These perturbations seeded nucleosynthesis and the first gravitational structure formation and should preserve information about the first turbulence before fossilization by inflation [6]. From HGD, the horizon scale $L_H = ct = 10^{20}$ m increases at light speed $c$ to match the Schwarz viscous scale $L_{SV} = (\gamma \nu / \rho G)^{1/2}$ at time $t = 10^{12}$ s (30,000 years) after the big bang, where $\gamma = 10^{-12}$ s$^{-1}$ is the rate of strain, $\rho_0 \sim 10^{-17}$ kg m$^{-3}$ and $\nu \sim 10^{26}$ m$^2$ s$^{-1}$ is the kinematic (photon) viscosity [2].

At $t = 10^{12}$ s protosupercluster mass gravitational structures form by fragmentation at density minima in the plasma epoch and expand at near sonic speed $V_s = c/3^{1/2}$ until the plasma cools to form gas at $t = 10^{13}$ s (30,000 years), as shown in Figure 1abc [7]. Supervoid regions up to 300 Mpc devoid of galaxies have been observed by radio telescopes, matching a cosmic microwave background (CMB) cold spot [8], and with 2-4-8-16 spherical harmonic axes aligned with spiral galaxy spin directions in the local 30 Mpc ($10^{24}$ m) supercluster region. The direction (RA=202°, δ=25°) on the celestial sphere [8] has been termed the "axis of evil" due to this unexpected violation of the cosmic principle that assumes no such preferred axis exists. From HGD the "axis of evil" is a form of fossil vorticity turbulence.

---

[1] Kinematic viscosity $\nu = L_C \text{v}$, where $L_C$ is the collision length of a particle moving with velocity $\text{v}$. Frictional momentum transfer by particle collisions gives Planck particles and big bang turbulence terminated by muon-viscosity inflation, photon-viscosity giving plasma protoclusters, protogalaxies and the second turbulence, gas viscosity giving starcluster planet clumps, planet-viscosity to stablize the clumps, and starclump-viscosity to keep galaxy clusters stable.
[2] Diffusivity $D = L_C \text{v}$ prevents any gravitational condensation or hierarchical clustering of collisionless CDM halos.
[3] Turbulence is defined as an eddy-like state of fluid motion where the inertial-vortex forces $\vec{\text{v}} \times \vec{\omega}$ of the eddies are larger than any other forces that tend to damp the eddies out. By this definition turbulence always cascades from small scales to large (contrary to the standard, but incorrect, assumption).
[4] Fossil turbulence is a perturbation in any hydrophysical field produced by turbulence that persists after the fluid is no longer turbulent at the scale of the perturbation.
[5] Fossil turbulence waves dominate the radial heat, mass, momentum, chemical species and information transport processes of stably stratified natural fluids such as stars in a beamed secondary (zombie) turbulence maser action.

Big bang turbulence Taylor microscale Reynolds numbers increased from critical $Re_\lambda \approx 10$ to strong turbulent $Re_\lambda \approx 1000$ values at strong force freeze-out, when muon-viscosity damped the turbulent combustion of Planck particles [4] and began exponential stretching of space by negative normal stresses according to Einstein's equations of general relativity [3]. Fossil temperature turbulence patterns seeded nucleosynthesis and the first gravitational structure formation, as shown in Figure 1e. Gravitational structure formation instability is nonlinear and absolute, so in the absence of fluid motions density maxima and density minima are equally unstable. However, the early universe was expanding rapidly with rate of strain $\gamma = t^{-1}$. Therefore the first structures to form were protosupervoids, by the mechanism illustrated in Figure 1e.

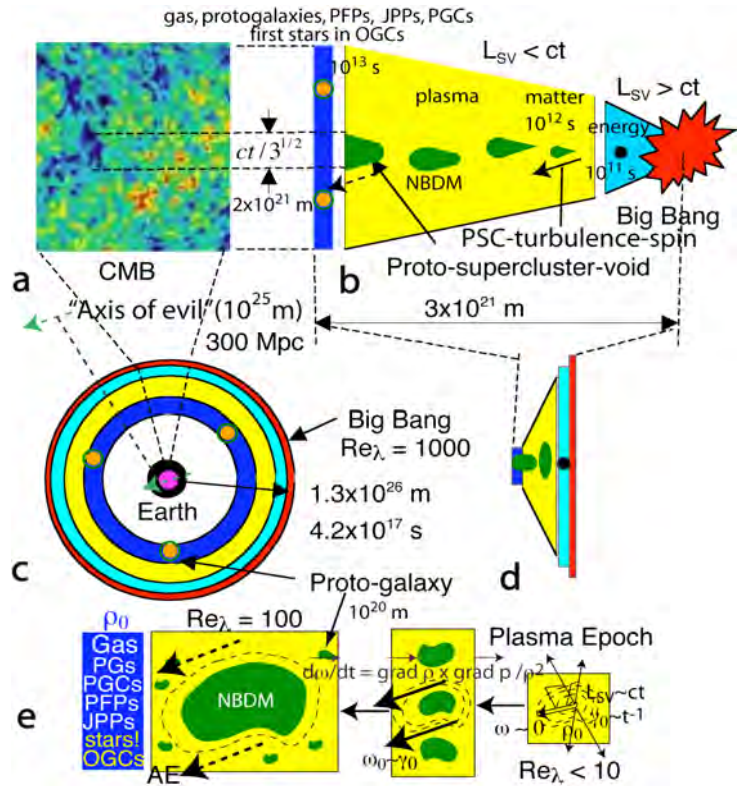

Fig. 1. HGD theory [1-7] predictions of the evolution of gravitational structure and the formation of supervoid and supercluster scale spins and the "Axis of Evil" [8]. Viscosity, turbulence, and fossil turbulence are important at every stage, contrary to standard collisionless ΛCDMHC models and numerical simulations [8,9]. Because $\nu$ decreases by a factor of $10^{13}$ at recombination, the fragmentation mass decreases to Earth-mass PFPs ($10^{25}$ kg) within million-solar-mass ($10^{36}$ kg) Jeans acoustic scale clumps (PGCs). The PGC clumps of PFP planets become increasingly diffusional as the planets freeze, forming the 0.3 Mpc BDM galaxy halo from the 0.003 Mpc protogalaxy core as shown by HST Tadpole images [10].

Figure 1e (right) shows a density minimum at transition to gravitational void formation. Pressure gradient and density gradient vectors need not be aligned, so vorticity will be produced at a rate $\partial\vec{\omega}/\partial t = grad\rho \times gradp/\rho^2$ on both sides of the low density region, leading to turbulence of the protosupercluster plasma remaining as the voids expand as rarefaction waves. The rate-of-strain and spin $\gamma_0 = \omega_0 = 10^{-12} \text{s}^{-1}$ and the plasma density $\rho_0 = 4\times 10^{-17} \text{kg m}^{-3}$ at $t = 10^{12}$ s are preserved as turbulent fossils of this time of first structure.

The baryonic supercluster mass $\rho_0 \times L_H^3 \approx 10^{46}$ kg at $t = 10^{12}$ s, approximately that observed, where $L_{SV} \approx L_H$, assuming the non-baryonic mass (green in Fig. 1) exceeds the baryonic mass by $10^{2-3}$. In the weakly turbulent plasma at $t = 10^{13}$ s, fragmentations are triggered along vortex lines at density minima just before the transition from plasma to gas at the $10^{20}$ m Kolmogorov scales $L_K \equiv (v^3/\varepsilon)^{1/4}$ with $L_K \approx L_{SV}$ and $L_{SV} \approx L_{ST}$, where $L_{ST} \equiv \varepsilon^{1/2}/(\rho G)^{3/4}$, $\varepsilon$ is the viscous dissipation rate and $G$ is Newton's gravitational constant, giving protovoids between protogalaxy gas clouds that expand limited by PGC-viscosity friction. These clouds should form linear clusters of protogalaxies along vortex lines with spiral clump-clusters of protogalaxies at the base of vortex lines following this universal Nomura-Post geometry of turbulence, demonstrated by direct numerical simulation of the Navier-Stokes equations for turbulence. The most common principle-rate-of-strain values for turbulence show stretching in two directions and compression in one [6], which explains observations of the HST ultra deep field observations that the second dimmest proto-galaxies have this spiral clump cluster geometry [11]. It is this geometry with proto-galaxies weakly stretching apart on the same plane that is most likely to produce Tadpole-like merger systems with a fragmented proto-galaxy core returning by gravitational attraction on the accretion disk of a companion.

## 2. Tadpole galaxy merger system

The Tadpole (VV29, UGC 10214) galaxy merger HST advanced camera for surveys (ACS) images provide clear evidence that dark matter galaxy halos are baryonic and frictional, as shown in Fig. 2. Suggestions that the Tadpole tail VV29b points to an invisible CDM halo [12] or is a Toomre & Toomre [13] frictionless tidal tail [14] are quite untenable in view of the wealth of detail in the HST/ACS images [10]. Spectroscopic studies with the Keck telescope [15] show the numerous globular star clusters in VV29b are young and were formed in place, justifying the label "star-wake" at the bottom of Fig. 1 (top) and confirming the Schild quasar microlensing interpretation [16] that the mass of galaxies is dominated by rogue planets that are the source of all stars from HGD. Some of the dark-matter PGCs in Tadpole have diffused out of the $L_N$ scale core in clumps that are revealed as a super-star-cluster SSC [15] labeled as a Dark Dwarf Galaxy in Fig. 1 (top). A dark halo boundary edge is shown by Fig. 3, and close-ups of star-wakes and dust-lanes justify identification of the merging $L_N$ scale fragmented proto-galaxy VV29cdef in Figs. 4-6.

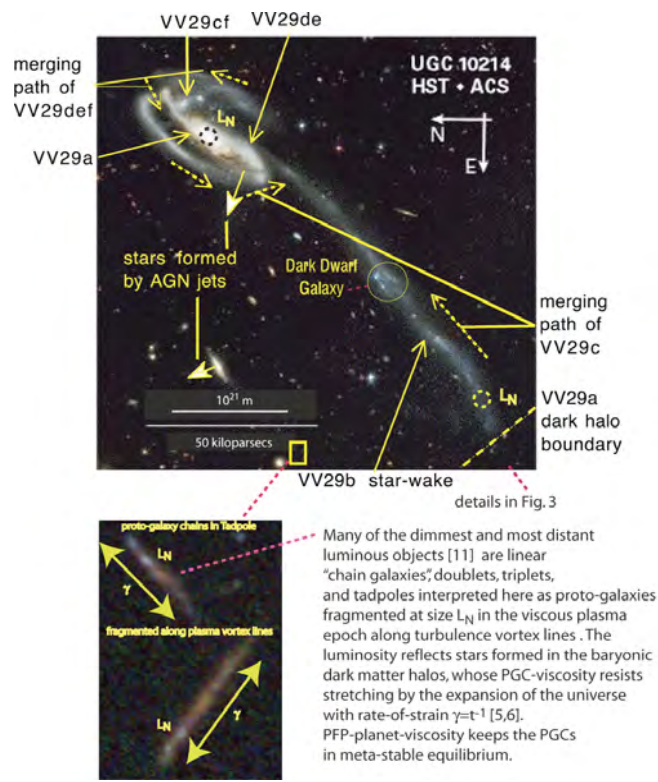

Fig. 2. Tadpole VV29 (UGC 10214) galaxy merger system. Galaxy fragments VV29cdef trigger star formation in the baryonic dark matter halo by tidal forces to produce the VV29b wake of young globular star clusters from planet clumps (top). Linear proto-galaxy clusters are seen in the enlarged box (bottom).

Dashed arrows in Fig. 1 (top) show the path of the VV29c galaxy and VV29f galaxy fragments as they spiral once around VV29a toward their position about $4L_N$ from the central core. Fragments VV29def are embedded in the VV29a disk. Both the embedded galaxy fragment VV29e and a background spiral galaxy have AGN jets (see Fig. 1 and Fig. 4) that trigger star formation in the BDM galaxy halos.

Two linear "chain galaxies" are shown in Fig. 1 (bottom), reflecting fragmentation of the weakly turbulent plasma before recombination to form proto-galaxies stretched on vortex lines by both the turbulence and the expansion of the universe. The dim luminosity between protogalaxies is interpreted as tidally agitated YGCs that provide frictional stresses that resist further separations and reduce the rate-of-strain $\gamma$ of the linear galaxy cluster toward zero.

Fig. 3 shows the VV29a BDM halo boundary edge at $40\,L_N$ from the central core, indicating the halo diameter is $0.3\,\text{Mpc}$. The diameter (dashed circle) of the star wake is $L_N$ suggesting tidal forces that triggered star formation from the PGC clumps of BDM planets were on the $L_N$ scale of a galaxy core fragment. The glow between the bright YGCs are field stars formed in the clumps and extracted by tidal forces as VV29cdef galaxy fragments move through the BDM halo.

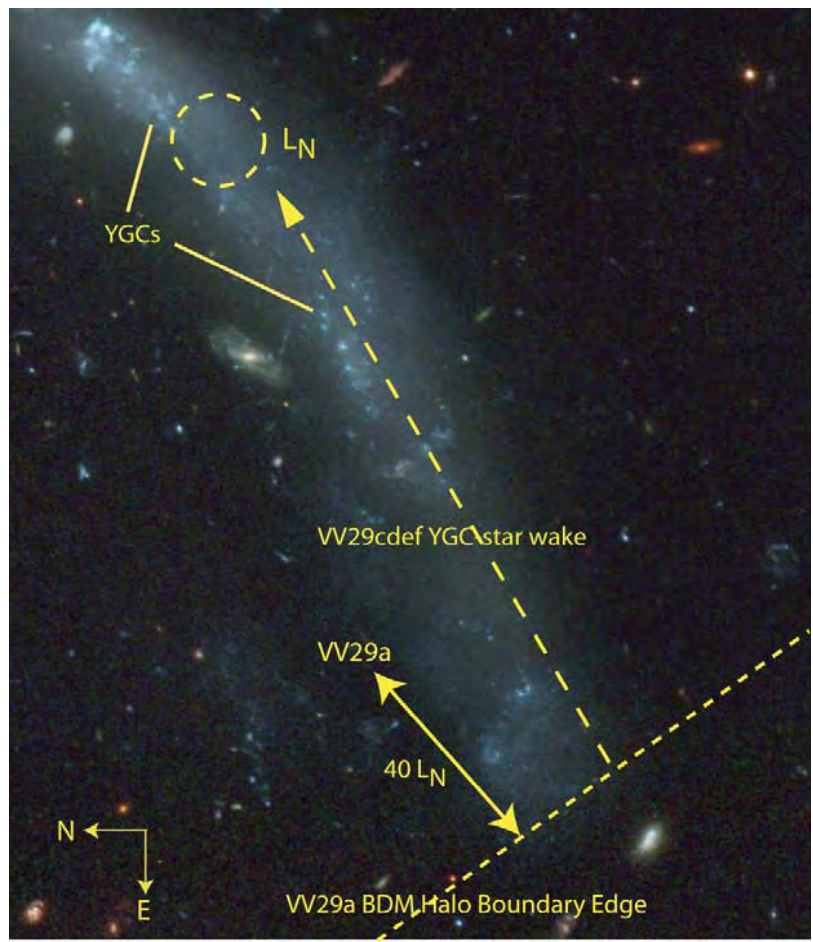

Fig. 3. Boundary of galaxy dark matter halo shown by a sharp beginning of the YGC-star wake.

Figure 4 shows a figure (top) from the Tran et al. Keck spectroscopic study of young globular clusters (YGCs) and a superstarcluster (SSC) in Tadpole [15]. Extrapolating the row of 46 YGCs matches the entry of the VV29c into its spiral capture by VV29a shown in Fig. 1. Fig. 4 (bottom) shows a background spiral galaxy with an AGN jet. The jet triggers star formation in the BDM galaxy halo, just as the jet from VV29e triggers star formation, as indicated in Fig. 1. The spiral galaxy PGC accretion disk appears to have tilted counterclockwise over time as shown by the arrows, leaving the curved trail of stars formed in its halo as a fossil. The mechanism is to increase the gas atmospheres and thus the accretion rates of the frozen H-He dark matter planets by tidal friction and radiation, increasing the formation rate of larger planets and stars.

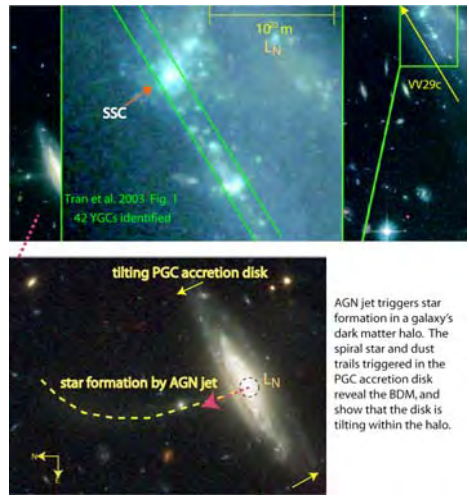

Fig. 4.  Young globular star clusters precisely in a line are triggered by the passage of VV29c (top).  Star formation is similarly produced by an active-galactive-nucleus (AGN) plasma jet into the BDM halo of a background galaxy (bottom and Fig. 1).

Figure 5 is a detailed image of the SSC identified in the Tran et al. study [15].  It appears to be a clump of PGCs in the BDM halo of VV29a, triggered to form stars in YGCs by the passage of VV29c.  From its size and brightness the dark/luminous mass is here estimated to be large ($10^3$), so it is labeled as a dark dwarf galaxy in Fig. 1 (top).  From the photometry [15], all the stars of the VV29b filamentary galaxy were formed in place in the halo, and therefore were not ejected as a frictionless Toomre & Toomre tidal tail [13].

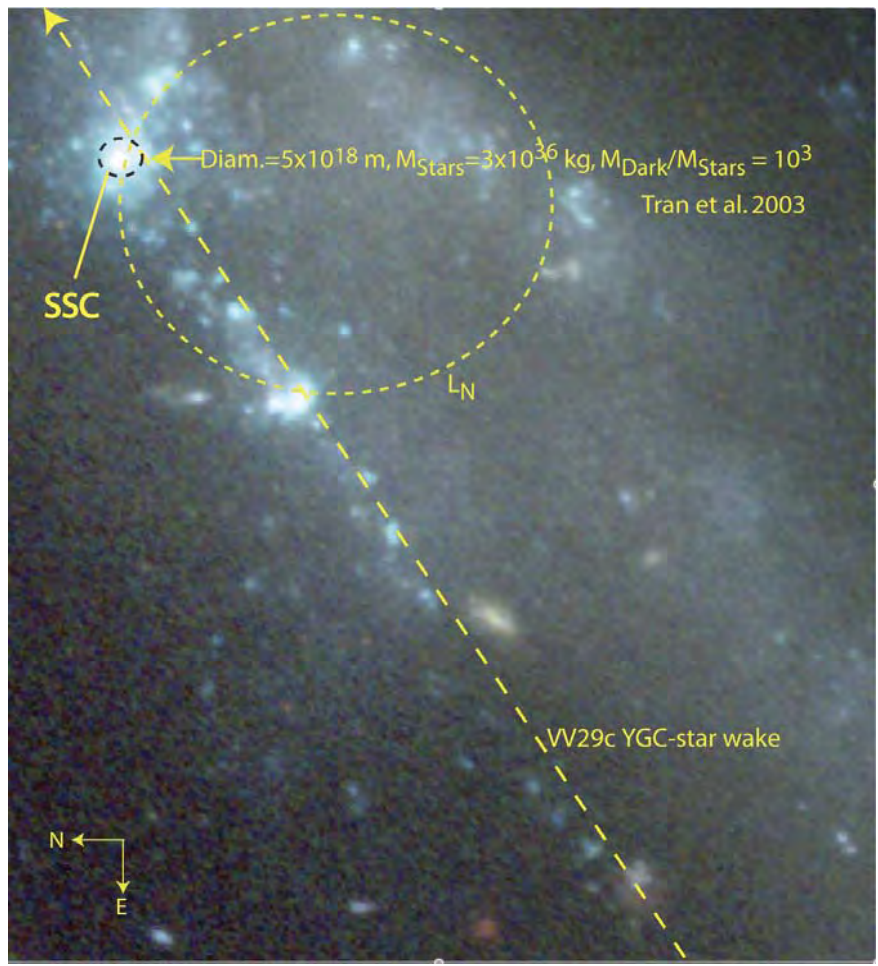

Fig. 5.  Detail of Tran et al. Superstar cluster and YGC-star wake.

Figure 6a shows the galaxy fragment VV29c finally embedded in the disk of VV29a after a frictional spiral star wake passage (left) completely around the VV29a core first below and then into its accretion disk. The star-wake and star-dust-wakes of fragments VV29def (center) are first above and then in the disk. Fig. 6b shows the Eastern part of the HST/ACS image. Note the curved VV29f dust wake above the VV29a disk.

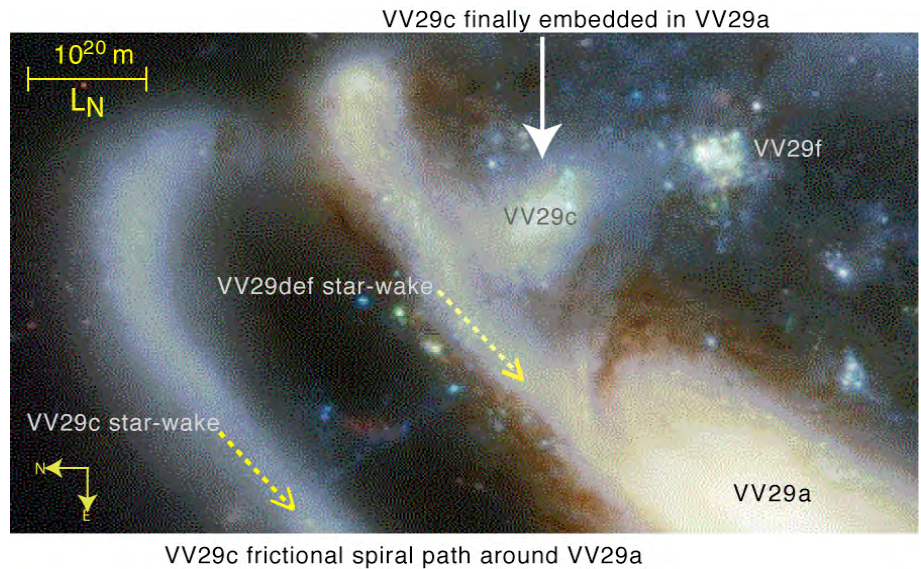

Fig. 6a. Western detail of HST/ACS Tadpole image showing frictional spiral star-wake and dust-wake captures of the galaxy fragments.

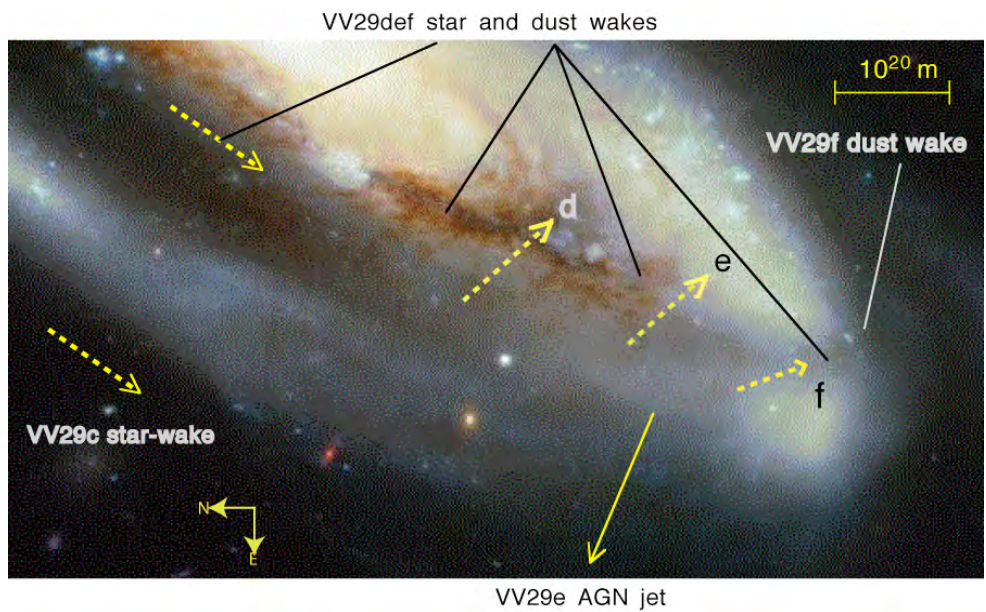

Fig. 6b. Eastern detail of the HST/ACS Tadpole image. The VV29a star-wake passes below the VV29a disk and the VV29f dust-wake passes above on their way to the embedded position shown in Fig. 5a. VV29d and VV29e fragments were captured within the disk. VV29e has an AGN jet that triggers star formation in the BDM halo of VV29a in the direction shown by the arrow.

## 3. Discussion

The HST/ACS images shown in Figs. 1-5 and Figs. 6ab support the Gibson 1996 HGD prediction [1] that the dark matter of galaxies should be frozen H-He rogue planets, as inferred independently by Schild 1996 from quasar microlensing [16]. The planets in their dense primordial clumps merge to form larger planets and all the stars. When the stars die they compress and spin, and spin-radiation evaporates the ambient planets to form planetary nebulae and a systematic dimming error. Widely accepted ΛCDMHC models of gravitational structure formation must therefore be abandoned along with all collisionless models of galactic dynamics [13, 17]. Claims that the universe expansion rate is accelerating due to an anti-gravity "dark energy" term in general relativity theory [18, 19, 20] are unnecessary because dark matter planets explain the random dimming of Supernovae Ia (SNe Ia) events as a systematic error not taken into account [7]. Estimates [21, 22] of Hubble Constant $H_0 = 1/T$ km s$^{-1}$Mpc$^{-1}$ values significantly smaller than those indicated by CMB data, suggesting the age of the universe T is 15.9 Gyr rather than 13.7 Gyr from the CMB observations of WMAP, and from HGD theory, suffer from the same dimming error from evaporated dark matter planet atmospheres, and are also unnecessary, as shown below in Fig. 7.

The Tadpole BDM halo diameter 0.3 Mpc is shown on the left of Fig. 7 [22], and can account for the observed "cool local flow" of galaxies due to PGC-viscous damping in the distance range 1-3 Mpc rather than dark energy [20, 23]. The Sandage et al. 2006 [21] estimates of the Hubble constant and age of the universe differ significantly from the WMAP and HGD values 71 km/s Mpc and 13.7 Gyr. The discrepancy is attributed to systematic errors from ambient BDM planet atmospheres surrounding white dwarfs that grow to Chandrasekhar instability and a SNe Ia event embedded in the Oort cavity produced when dark matter planets in a PGC accrete to form a star [7]. SNe Ia events are thus dimmed, or not, by planet atmospheres. Those that are not dimmed correspond to the dashed (13.7 Gyr) line of Fig. 7.

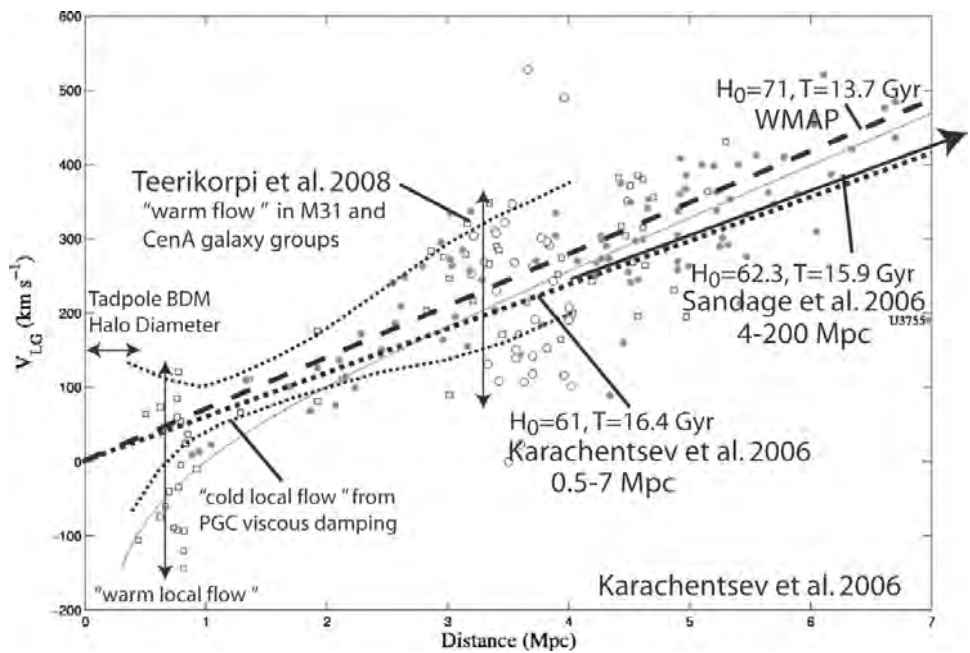

Fig. 7. Estimates of the Hubble Constant $H_0$ = km/s Mpc from redshift velocities versus distance [21, 22]. A systematic error in the SNe Ia distance estimates due to BDM planet dimming of the SNe Ia brightnesses [7] accounts for the overestimates of the age of the universe compared to the WMAP value of 13.7 Gyr.

On the left of Fig. 7 under the dominant influences of PGC viscosity and gravity the local group galaxies interact much as the molecules of gas in thermal equilibrium. The dotted "cold local flow" envelope for distances 1-3 Mpc are interpreted as those galaxies whose speeds have been damped to near zero by PGC viscous damping so that the Hubble flow expansion of the universe is their dominant influence so they can be convected out of the local group by Hubble flow. The scatter about the WMAP Hubble flow line beginning at 3 Mpc is interpreted as the PGC viscous "virialization" of the M31 and the CenA groups of galaxies by the same mechanism [24, 25].

## 4. Conclusions

High resolution HST/ACS images of the Tadpole galaxy merger system show the merger is frictional, where the source of friction is evaporating dark-matter-planets (PFPs) forming larger planets and stars. Tidal interactions occur between the VV29a galaxy dark matter halo of planet clumps (PGCs) and merging VV29cdef galaxy fragments that pass through the VV29a halo leaving young globular star cluster (YGC), star-dust, and field-star wakes on spiral paths before the fragments eventually embed themselves in the inner VV29a spiral galaxy disk. From the spacing and mass of the YGCs and the 0.3 Mpc size of the halo it appears that most of the VV29a galaxy mass exists in the dark matter halo, which from HGD theory diffused out of the protogalaxy core formed in the plasma epoch after fragmentation at recombination at planet (PFP) and protoglobularstarcluster (PGC) masses and freezing to solid H-He dark objects as the expanding universe cooled, Fig. 1e.

We see from Tadpole that stars form from planets in dense clumps of planets, not from gas clouds falling into merging, mythical, CDM halos. The first stars formed gently at 0.3 Myr from white-hot planet-mass clouds soon after plasma-gas recombination to form present day old globular clusters (OGCs), lighting up the linear protogalaxy clusters of Fig. 2 (bottom). There were no dark ages before 300 Myr without stars or light. Population III superstars never happened, and neither did re-ionization. The planets of PGCs are held in metastable equilibrium by PFP-viscosity. Protogalaxies were pre-formed in protoclusters during the plasma epoch and resist separation by the expansion of the universe by PGC-viscosity, Fig. 7. Galaxy dark matter halos form from PGC-diffusivity and are mostly baryonic due to large NBDM-diffusivity values forming low density but massive galaxy-cluster halos. Supervoids observed extending to 300 Mpc originated at 0.03 Myr and produced 30 Mpc supercluster turbulence and spiral galaxy spin correlations [8].

Dark-energy cold-dark-matter hierarchical-clustering ($\Lambda$CDMHC) collisionless models [17, 18, 19] for the formation and evolution of gravitational structures in the universe are inconsistent with the Tadpole images, other data and HGD theory [1-10, 16]. Frictionless galactic bridges and tails [13] are rendered obsolete and misleading by the Tadpole images.